\documentclass[conference,a4paper]{IEEEtran}
\usepackage{cite}
\usepackage{bm}
\usepackage{float}
\usepackage{array}
\usepackage{multirow}
\usepackage[cmex10]{amsmath}
\usepackage{graphicx}
\usepackage{url}
\usepackage{color}
\usepackage{amssymb}
\usepackage{amsfonts}
\usepackage{amsmath}
\usepackage{extarrows}
\usepackage{graphicx}
\usepackage{amsfonts}
\usepackage{caption}
\usepackage{algorithm}
\usepackage{algpseudocode}
\usepackage{indentfirst}
\usepackage{caption}
\usepackage{esint} 
\usepackage{graphicx}
\usepackage{graphics}
\usepackage{pgfplots}
\usepackage{tikz}
\usepackage{epsfig}
\usepackage{soul}
\usepackage{caption}
\usepackage{gensymb}
\def\bA{{\mathbf{A}}} \def\bB{{\mathbf{B}}}  \def\bD{{\mathbf{D}}} 
 \def\bG{{\mathbf{G}}} \def\bH{{\mathbf{H}}} \def\bI{{\mathbf{I}}} \def\bJ{{\mathbf{J}}}
   \def\bN{{\mathbf{N}}} 
\def\bP{{\mathbf{P}}}    \def\bT{{\mathbf{T}}}
    \def\bY{{\mathbf{Y}}}

\def\ba{{\mathbf{a}}}    
  \def\bh{{\mathbf{h}}}  
   \def\bn{{\mathbf{n}}} 
\def\bp{{\mathbf{p}}}    
   \def\bx{{\mathbf{x}}} \def\by{{\mathbf{y}}}

\begin{document}

\title{Simultaneous Indoor and Outdoor 3D Localization with STAR-RIS-Assisted Millimeter Wave Systems}
\author{\IEEEauthorblockN{Jiguang~He$^1$,
Aymen~Fakhreddine$^1$, and George C. Alexandropoulos$^{1,2}$}
\IEEEauthorblockA{$^1$Technology Innovation Institute, 9639 Masdar City, Abu Dhabi, United Arab Emirates}
\IEEEauthorblockA{$^2$Department of Informatics and Telecommunications, National and Kapodistrian University of Athens\\
Panepistimiopolis Ilissia, 15784 Athens, Greece}
\IEEEauthorblockA{e-mails: \{jiguang.he, aymen.fakhreddine\}@tii.ae, alexandg@di.uoa.gr}
}
\maketitle
\thispagestyle{plain}
\pagestyle{plain}

\begin{abstract}
Simultaneously transmitting (refracting) and reflecting reconfigurable intelligent surfaces (STAR-RISs) have been recently identified to improve the spectrum/energy efficiency and extend the communication range. However, their potential for enhanced concurrent indoor and outdoor localization has not yet been explored. In this paper, we study the fundamental limits, i.e., the Cram\'er Rao lower bounds (CRLBs) via Fisher information analyses, on the three-dimensional (3D) localization performance with a STAR-RIS at millimeter wave frequencies. The effect of the power splitting between refraction and reflection at the STAR-RIS as well as the power allocation between the two mobile stations (MSs) are investigated. By maximizing the principal angle between the two subspaces corresponding to the STAR-RIS reflection and refraction matrices, we are able to find the optimal solutions for these simultaneous operations. We verify that high-accuracy 3D localization can be achieved for both indoor and outdoor MSs when the system parameters are well optimized.   
\end{abstract}

\section{Introduction}
Reconfigurable intelligent surfaces (RISs) have recently been introduced for improved energy efficiency (EE), spectrum efficiency (SE), localization accuracy, sensing capability, as well as network/physical-layer security~\cite{huang2019reconfigurable, risTUTORIAL2020, wymeersch2019radio,hu2020reconfigurable,RISE6G_COMMAG}. The RIS, either being passive, active, or hybrid, is used to smartly control the radio propagation environment, by virtue of multi-function capabilities, e.g., reflection, refraction, diffraction, polarization, scattering, and even absorption~\cite{di2019smart,huang2019reconfigurable}. In the literature, the RIS is commonly used as an intelligent reflector, which breaks the well-known law of reflection~\cite{Alexandropoulos2020}, to mitigate the blockage and shadowing effect, and expand the communication coverage, especially for millimeter wave (mmWave) and Terahertz (THz) communications. 

Recently, there appears another new type of RIS, termed as simultaneously transmitting and reflecting RIS (STAR-RIS)~\cite{Yuanwei2021}, which can be regarded as a promising bridge linking indoor and outdoor connectivity. Unlike the conventional RIS, the STAR-RIS can simultaneously realize two different functionalities, e.g., reflection and refraction. Namely, mobile stations (MSs) on both sides of the STAR-RIS, for instance, one is indoor and the other is outdoor, can be served by it at the same time. In addition, the two functionalities can also be realized in different manners, e.g., energy splitting, time switching, and model switching.   

Even though the potential of traditional reflective RISs for radio localization has been studied in the literature~\cite{Jiguang2020,Elzanaty2021}, we envision that the localization capability can be even further extended by the introduction of one or multiple STAR-RISs. With the aid of such a surfaces, an outdoor base station (BS) is capable of localizing an indoor user, in addition to an outdoor user, via uplink sounding reference signal (SRS) transmission, not only at sub-6GHz but also mmWave frequency bands. In this paper, we study the Cram\'er Rao lower bounds (CRLBs) for the intermediate channel parameter estimation based on the Fisher information analysis, and extend it to three-dimensional (3D) localization of one indoor and one outdoor MSs by virtue of the Jacobian matrix. We also examine the effect of the energy splitting between the two STAR-RIS functionalities and power allocation between the two MSs during SRS transmission. Besides, we find the optimal design of the STAR-RIS based on the CRLBs, which offers better performance than other alternative designs.

\textit{Notations}: A bold lowercase letter $\ba$ denotes a vector, and a bold capital letter $\bA$ denotes a matrix. $(\cdot)^\mathsf{T}$ and $(\cdot)^\mathsf{H}$ denote the matrix or vector transpose and Hermitian transpose, respectively. $(\cdot)^{-1}$ denotes inverse of a matrix, $\mathrm{tr}(\cdot)$ denotes the trace operator, $\mathrm{diag}(\ba)$ denotes a square diagonal matrix with the entries of $\ba$ on its diagonal, 
$\bA \otimes \bB$ and $\bA \diamond \bB$ denote the Kronecker and Khatri-Rao products of $\bA$ and $\bB$, respectively,
$\mathbb{E}[\cdot]$ and $\mathrm{var}(\cdot)$ are the expectation and variance operators, $\mathbf{1}$ is the all-one vector, $\bI_{M}$ denotes the $M\times M$ identity matrix, $j = \sqrt{-1}$, and $\|\cdot\|_2$ denotes the Euclidean norm of a vector. $[\ba]_i$, $[\bA]_{ij}$, and $[\bA]_{i:j, i:j}$ denote the $i$th element of $\ba$, the $(i,j)$th element of $\bA$, and the submatrix of $\bA$ formed by rows $i,i+1, \ldots, j$ and columns $i,i+1, \ldots, j$. Finally, $|\cdot|$ returns the absolute value of a complex number. 

\section{System Model}
\begin{figure}[t]
	\centering
\includegraphics[width=1\linewidth]{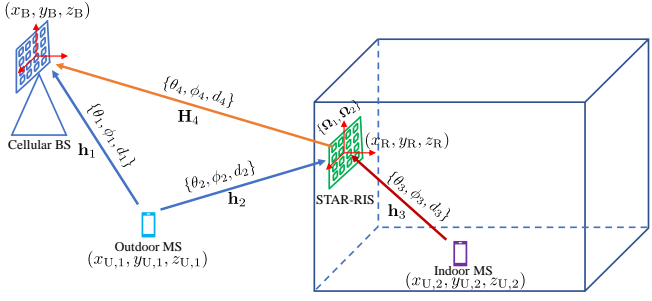}
	\caption{The considered STAR-RIS-assisted mmWave MIMO systems for simultaneous indoor and outdoor 3D localization.}
		\label{System_Model}
		\vspace{-0.5cm}
\end{figure}
We consider a nearly-passive STAR-RIS-aided mmWave multiple-input multiple-output (MIMO) system, which consists of one multi-antenna BS, one multi-element STAR-RIS, one indoor MS, and one outdoor MS, which is intended for simultaneous indoor and outdoor 3D localization, as shown in Fig.~\ref{System_Model}. The BS and STAR-RIS are equipped with $M$ antennas and $N$ almost passive scattering elements, respectively, while each MS is equipped with a single antenna. Extension to multi-antenna MSs is feasible. Both the BS and STAR-RIS employ the uniform planar array (UPA) structure parallel to the $x-z$ plane. The STAR-RIS has two operation functionalities, i.e., reflection and refraction, which can be realized simultaneously. 

\subsection{Channel Model}
The system is assumed to operate in the mmWave frequency band and we consider the 
Saleh-Valenzuela parametric channel model to construct all four individual channels. The direct line-of-sight (LoS) channel between the outdoor MS and the $M$-antenna cellular BS is denoted as $\bh_1\in\mathbb{C}^{M \times 1}$ and is mathematically expressed as follows:
\begin{equation}\label{h_1}
\bh_1 = \frac{e^{-j 2\pi d_1/\lambda}}{\sqrt{\rho_1}} \boldsymbol{\alpha}_x(\theta_{1},\phi_{1}) \otimes \boldsymbol{\alpha}_z(\phi_{1}),
\end{equation}
 where $d_1$ (in meters) and $\rho_1$ (for the sake of simplicity, we assume that $\rho_1 = d_1^2$) are the distance and path loss between the outdoor MS and BS, respectively, $\lambda$ is the wavelength of the carrier frequency, $\theta_1$ and $\phi_1$ are the azimuth and elevation angles of arrival associated with $\bh_1$, respectively.\footnote{In fact, we can also consider the free-space path loss, which is modeled: 
\begin{equation}
    \rho_1 = d_1^2 f_c^2 / 10^{8.755}, \nonumber
\end{equation}
where $f_c$ (in KHz) is the carrier frequency, defined as $f_c = \frac{c}{\lambda}$ with $c$ being the speed of light ($3\times 10^8$ m/s). Alternatively, the standard 3GPP urban micro (UMi) path
loss model can be considered, according to which holds: 
\begin{equation}
    \rho_1 = 10^{2.27} d_1^{3.67} f_c^{2.6}, \nonumber
\end{equation}
where $f_c$ needs to be included in GHz~\cite{Akdeniz2014}.
} 
The steering vectors $\boldsymbol{\alpha}_x(\theta_{1},\phi_{1})$ and $\boldsymbol{\alpha}_z(\phi_{1})$ can be written as~\cite{Tsai2018}:
 \begin{align}
     \boldsymbol{\alpha}_x(\theta_{1},\phi_{1}) =& \Big[e^{-j \frac{2\pi d_x}{\lambda} (\frac{M_x -1}{2}) \cos(\theta_1) \sin(\phi_1)},\nonumber \\
    & \cdots, e^{j \frac{2\pi d_x}{\lambda} (\frac{M_x -1}{2}) \cos(\theta_1)\sin(\phi_1)} \Big]^{\mathsf{T}},\\
       \boldsymbol{\alpha}_z(\phi_{1}) =& \Big[e^{-j \frac{2\pi d_z}{\lambda} (\frac{M_z -1}{2}) \cos(\phi_1) },\nonumber \\
    & \cdots, e^{j \frac{2\pi d_z}{\lambda} (\frac{M_z -1}{2}) \cos(\phi_1)} \Big]^{\mathsf{T}},
 \end{align}
where $M = M_x M_z$ with $M_x$ and $M_z$ being the number of horizontal and vertical BS antennas, respectively, and $d_x$ and $d_z$ denote their inter-element spacing in the horizontal and vertical axes, which were set as half-wavelength without loss of generality. Similarly, the other channels, e.g., $\bh_2 \in \mathbb{C}^{N \times 1}$ and $\bh_3\in \mathbb{C}^{N \times 1}$, can be presented in the same manner, as: 
 \begin{equation}\label{h_2_h_3}
\bh_i = \frac{e^{-j2\pi d_i/\lambda}}{\sqrt{\rho_i}} \boldsymbol{\alpha}_x(\theta_{i},\phi_{i}) \otimes \boldsymbol{\alpha}_z(\phi_{i}),
\end{equation}
for $i = 2$ and $3$, where $N = N_x N_z$ is the number of STAR-RIS elements with $N_x$ and $N_z$ denoting the numbers in the horizontal and vertical axes, respectively. Note that the array response vectors $\boldsymbol{\alpha}_x(\cdot)$ and $\boldsymbol{\alpha}_z(\cdot)$ may have a different dimension compared to that in~\eqref{h_1} but possess the same form.  Finally, $\rho_2$ and $\rho_3$ follow the same assumption as $\rho_1$. 
 
The wireless channel between the STAR-RIS and BS, i.e., $\bH_4 \in \mathbb{C}^{M \times N}$, is expressed as follows:
 \begin{equation}\label{H_4}
\bH_4 = \frac{e^{-j 2\pi d_4/\lambda}}{\sqrt{\rho_4}} \boldsymbol{\alpha}_x(\theta_{4},\phi_{4}) \otimes \boldsymbol{\alpha}_z(\phi_{4})(\boldsymbol{\alpha}_x(\theta_{4},\phi_{4}) \otimes \boldsymbol{\alpha}_z(\phi_{4}))^\mathsf{H},
\end{equation}
assuming that the two UPAs (i.e., one for the BS and the other for the STAR-RIS) are deployed in parallel without any biased orientation.  Due to the wall in between the BS and the indoor MS, there does not exist a direct LoS path between them. The only path is the reflection route via the STAR-RIS. In this work, we consider only LoS paths in all the individual channels; the extension for the multipath scenario and for arbitrary orientations between the UPAs is left for future work.

\subsection{Geometric Relationship}
The Cartesian coordinates of the BS and STAR-RIS as well as the outdoor and indoor MSs are $\bp_\text{B} = (x_\text{B},y_\text{B},z_\text{B})^\mathsf{T}$, $\bp_\text{R} = (x_\text{R},y_\text{R},z_\text{R})^\mathsf{T}$, $\bp_{\text{U},1} = (x_{\text{U},1},y_{\text{U},1},z_{\text{U},1})^\mathsf{T}$, and $\bp_{\text{U},2} = (x_{\text{U},2},y_{\text{U},2},z_{\text{U},2})^\mathsf{T}$, respectively. The relationship between the distances and a pair of Cartesian coordinates are listed below:
\begin{align}
    d_1 &= \|\bp_\text{B} - \bp_{\text{U},1} \|_2, \\
     d_i &=  \|\bp_\text{R} - \bp_{\text{U},i-1} \|_2,\;\text{for}\; i = 2,3, \\
       d_4 &=  \|\bp_\text{B} - \bp_{\text{R}} \|_2.
\end{align}
By introducing the three-element vector $\boldsymbol{\xi}_i \triangleq [\cos(\theta_i) \cos(\phi_i), \sin(\theta_i) \cos(\phi_i), \sin(\phi_i) ]^\mathsf{T}$ for $i = 1,2,3,4$, the geometric relationship between the angular parameters and the Cartesian coordinates of the nodes can be expressed as 
\begin{align}\label{Geometry}
    \bp_{\text{R}} &= \bp_{\text{B}} + d_4 \boldsymbol{\xi}_4,  \\  
    \bp_{\text{U},1} & = \bp_{\text{B}} + d_1 \boldsymbol{\xi}_1 = \bp_{\text{R}} + d_2 \boldsymbol{\xi}_2,  \label{p_u_1_1} \\
     \bp_{\text{U},2} & = \bp_{\text{R}} + d_3 \boldsymbol{\xi}_3. \label{p_u_2}
\end{align}
 \subsection{Signal Model}
 Recall that the STAR-RIS has two operation functionalities, i.e., reflection and refraction. Two separate series of phase shifters are leveraged for controlling them, so each one is represented by a phase control matrix, i.e., $\boldsymbol{\Omega}_1$ for controlling refraction and $\boldsymbol{\Omega}_2$ for controlling reflection. $\boldsymbol{\Omega}_1$ and $\boldsymbol{\Omega}_2$ are diagonal matrices with each diagonal element satisfying the unit-modulus constraints, e.g., $|[\boldsymbol{\Omega}_1]_{jj}| = |[\boldsymbol{\Omega}_2]_{jj}| = 1$, $\forall j=1,2,\ldots,N$. We consider the 3D localization via the uplink transmission, where the two users send SRSs towards the BS simultaneously. 
The received signal during the $k$th time slot, for $k =1,2,\ldots,K$, can be mathematically expressed as
 \begin{equation} \label{by_k}
     \by_k = \bh_1 x_{1,k} + \epsilon_2 \bH_4\boldsymbol{\Omega}_{2,k}\bh_2 x_{1,k} + \epsilon_1 \bH_4\boldsymbol{\Omega}_{1,k}\bh_3 x_{2,k} + \bn_k,
 \end{equation}
where $x_{1,k}$ is the SRS from the outdoor MS, $x_{2,k}$ is the SRS from the indoor MS, as well as $\epsilon_1$ (for refraction) and $\epsilon_2$ (for reflection) are used to control the energy splitting for the two different operational modes of the STAR-RIS, satisfying the following condition: $\epsilon_1^2 + \epsilon_2^2 = 1$. The received signal at the BS is further corrupted by the white Gaussian noise $\bn_k$, and each element of $\bn_k$ follows $\mathcal{CN}(0, \sigma^2)$. During the $k$th time slot, the refraction matrix $\boldsymbol{\Omega}_{1,k}$ and the reflection matrix $\boldsymbol{\Omega}_{2,k}$ are considered. In order to ensure good estimates, $\boldsymbol{\Omega}_{1,k}$ and $\boldsymbol{\Omega}_{2,k}$ vary from one time slot to another, i.e., $\boldsymbol{\Omega}_{1,1} \neq \boldsymbol{\Omega}_{1,2} \neq \cdots \neq  \boldsymbol{\Omega}_{1,K}$, and $\boldsymbol{\Omega}_{2,1} \neq \boldsymbol{\Omega}_{2,2} \neq \cdots \neq  \boldsymbol{\Omega}_{2,K}$. This refractive/reflective beam sweeping was verified through our numerical results on the STAR-RIS design in Section~\ref{STAR_RIS_Design}.  
 
Based on the received signals across $K$ time slots, the BS estimates the Cartesian coordinates of both the indoor and outdoor users, enabling 3D localization. Without loss of generality, we assume that the sum power constraint is applied for each time slot, i.e., $\mathbb{E}[|x_{1,k}|^2] + \mathbb{E}[|x_{1,k}|^2] = P$, $\forall k$. The vector $\by_k$ in \eqref{by_k} can be further expressed as 
\begin{align}
    \by_k = &\bh_1 x_{1,k} + \epsilon_2 \bH_4 \mathrm{diag}(\bh_2) \boldsymbol{\omega}_{2,k} x_{1,k} \nonumber\\
    &+ \epsilon_1 \bH_4 \mathrm{diag}(\bh_3)\boldsymbol{\omega}_{1,k} x_{2,k} + \bn_k,
\end{align}
where $\boldsymbol{\Omega}_{1,k} = \mathrm{diag} (\boldsymbol{\omega}_{1,k})$ and  $\boldsymbol{\Omega}_{2,k} = \mathrm{diag} (\boldsymbol{\omega}_{2,k})$, $\forall k$. By stacking all $\by_k$'s column by column, we get the expression:
 \begin{align} \label{bY}
     \bY =& \eta_1 \sqrt{P} \bh_1 \mathbf{1}^\mathsf{T} + \eta_1 \sqrt{P} \epsilon_2 \bH_4 \mathrm{diag}(\bh_2) \bar{\boldsymbol{\Omega}}_2 \nonumber \\
     &+ \eta_2 \sqrt{P} \epsilon_1 \bH_4 \mathrm{diag}(\bh_3)\bar{\boldsymbol{\Omega}}_1 + \bN,
 \end{align}
where $\mathbf{1}$ denotes the $K$-element all-one vector, $\bY = [\by_1, \cdots, \by_K]$, $\bN = [\bn_1, \cdots, \bn_K]$, $\bar{\boldsymbol{\Omega}}_1 = [\boldsymbol{\omega}_{1,1}, \cdots, \boldsymbol{\omega}_{1,K}]$, and $\bar{\boldsymbol{\Omega}}_2 = [\boldsymbol{\omega}_{2,1}, \cdots, \boldsymbol{\omega}_{2,K}]$. We have also set $|x_{1,k}|^2 = \eta_1^2 P$,  $|x_{2,k}|^2 = \eta_2^2 P$, where $\eta_1^2 + \eta_2^2 =1$. Applying vectorization to $\bY$ in~\eqref{bY}, the following expression is deduced:
 \begin{align}\label{vec_Y}
     \by =& \eta_1\sqrt{P} (\mathbf{1} \otimes \bI_{M}) \bh_1+ \eta_1 \sqrt{P} \epsilon_2 (\bar{\boldsymbol{\Omega}}_2^\mathsf{T} \otimes \bI_M)(\bI_N \diamond \bH_4) \bh_2\nonumber\\
     &+ \eta_2 \sqrt{P} \epsilon_1 (\bar{\boldsymbol{\Omega}}_1^\mathsf{T} \otimes \bI_M)(\bI_N \diamond \bH_4) \bh_3 + \bn,
 \end{align}
 where $\by = \mathrm{vec}(\bY)$ and $\bn = \mathrm{vec}(\bN)$. 
We further introduce the following new notations to simplify the analyses that follows:
\begin{align}
    \bA_1 &=  (\mathbf{1} \otimes \bI_M), \\
    \bA_2 &=   (\bar{\boldsymbol{\Omega}}_2^\mathsf{T} \otimes \bI_M)(\bI_N \diamond \bH_4), \\
    \bA_3 & =    (\bar{\boldsymbol{\Omega}}_1^\mathsf{T} \otimes \bI_M)(\bI_N \diamond \bH_4).
\end{align}
To this end, expression in~\eqref{vec_Y} can be re-written as: 
 \begin{equation} \label{vec_Y1}
     \by =\sqrt{P} \eta_1\bA_1 \bh_1 +  \sqrt{P}\eta_1 \epsilon_2 \bA_2 \bh_2 + \sqrt{P}\eta_2\epsilon_1\bA_3 \bh_3 + \bn.
 \end{equation}
 Upon the STAR-RIS deployment, we assume that the BS knows the exact/precise location of the STAR-RIS. Thus, we assume that the BS has exact information on $\bH_4$ in terms of the parameters $\theta_4$, $\phi_4$, and $d_4$.\footnote{The assumption needs to be further relaxed since perfect information is usually infeasible in practice.} Therefore, $\bA_1$, $\bA_2$, and $\bA_3$ in~\eqref{vec_Y1} are known measurement matrices to the BS (the BS also knows the refractive/reflective phase configurations due to its interaction with the STAR-RIS controller).   
 
 \section{Cram\'er Rao Lower Bound Analyses}
In this section, we will provide the CRLBs on the estimation of the intermediate channel parameters, followed by the 3D Cartesian coordinates' estimation. We also present the refractive/reflective optimization of the STAR-RIS for the indoor/outdoor localization objective. 
 
\subsection{Estimation of Channel Parameters}
The channel parameters to be estimated are those included in $\bh_1$, $\bh_2$, and $\bh_3$, i.e., the nine-tuple $\boldsymbol{\nu} \triangleq [\theta_1, \phi_1, d_1,\theta_2, \phi_2, d_2,\theta_3, \phi_3, d_3]^\mathsf{T}$. Since the additive noise is complex Gaussian distributed, by introducing $\boldsymbol{\mu}(\boldsymbol{\nu}) \triangleq  \eta_1\bA_1 \bh_1 +  \eta_1 \epsilon_2 \bA_2 \bh_2 + \eta_2\epsilon_1\bA_3 \bh_3$ for~\eqref{vec_Y1}, the Fisher information matrix for $\boldsymbol{\nu}$ is obtained as:
\begin{equation}\label{Fisher_parameter}
[\bJ(\boldsymbol{\nu}) ]_{i,j} =  \frac{P}{\sigma^2}\Re \Big\{\frac{\partial \boldsymbol{\mu}^\mathsf{H}} {\partial \nu_i} \frac{ \partial \boldsymbol{\mu}}{ \partial \nu_j} \Big\}.
\end{equation}
The detailed information on the partial derivatives in~\eqref{Fisher_parameter} related to parameters in $\bh_1$, $\bh_2$, and $\bh_3$ is provided below: 
\begin{align}
    \frac{\partial \boldsymbol{\mu}} {\partial \theta_1} &= \eta_1\bA_1 (\bD_{11} \otimes \bI_{M_x})\bh_1,\\
     \frac{\partial \boldsymbol{\mu}} {\partial \phi_1} &= \eta_1\bA_1 (\bD_{12} \otimes \bI_{M_x})\bh_1 + \eta_1\bA_1 (\bI_{M_z} \otimes \bD_{13}) \bh_1, \nonumber \\
       & = \eta_1 \bA_1 (\bD_{12} \otimes \bI_{M_x}  + \bI_{M_z} \otimes \bD_{13})\bh_1, \\
    \frac{\partial \boldsymbol{\mu}} {\partial d_1} & = \eta_1 \frac{-j2\pi d_1/\lambda -1}{d_1} \bA_1 \bh_1,\\
    \frac{\partial \boldsymbol{\mu}} {\partial \theta_2} &= \eta_1\epsilon_2 \bA_2 (\bD_{21} \otimes \bI_{N_x})\bh_2,\\
     \frac{\partial \boldsymbol{\mu}} {\partial \phi_2} &= \eta_1\epsilon_2\bA_2 (\bD_{22} \otimes \bI_{N_x})\bh_2 + \eta_1\epsilon_2 \bA_2 (\bI_{N_z} \otimes \bD_{23}) \bh_2, \nonumber \\
       & = \eta_1 \epsilon_2\bA_2 (\bD_{22} \otimes \bI_{N_x}  + \bI_{N_z} \otimes \bD_{23})\bh_2, \\
    \frac{\partial \boldsymbol{\mu}} {\partial d_2} & = \eta_1\epsilon_2 \frac{-j2\pi d_2/\lambda -1}{d_2} \bA_2 \bh_2,\\
       \frac{\partial \boldsymbol{\mu}}
    {\partial \theta_3} &= \eta_2\epsilon_1 \bA_3 (\bD_{31} \otimes \bI_{N_x})\bh_3,\\
     \frac{\partial \boldsymbol{\mu}} {\partial \phi_3} &= \eta_2\epsilon_1\bA_3 (\bD_{32} \otimes \bI_{N_x})\bh_3 + \eta_2\epsilon_1 \bA_3 (\bI_{N_z} \otimes \bD_{33}) \bh_3, \nonumber \\
       & = \eta_2 \epsilon_1\bA_3 (\bD_{32} \otimes \bI_{N_x}  + \bI_{N_z} \otimes \bD_{33})\bh_3, \\
    \frac{\partial \boldsymbol{\mu}} {\partial d_3} & = \eta_2\epsilon_1 \frac{-j2\pi d_3/\lambda -1}{d_3} \bA_3 \bh_3,
\end{align}
where we have used the following matrix definitions:
\begin{align}
\bD_{11} =& \mathrm{diag}\Big(\Big[-j\pi \Big(-\frac{M_x -1}{2}\Big) \sin(\theta_1) \sin(\phi_1), \cdots, \nonumber\\ & -j\pi \Big(\frac{M_x -1}{2}\Big) \sin(\theta_1) \sin(\phi_1)\Big]\Big),
\end{align}
\begin{align}
\bD_{12} =& \mathrm{diag}\Big(\Big[j\pi \Big(-\frac{M_x -1}{2}\Big) \cos(\theta_1) \cos(\phi_1), \cdots,\nonumber\\ &j\pi \Big(\frac{M_x -1}{2}\Big) \cos(\theta_1) \cos(\phi_1)\Big]\Big),\\
\bD_{13} &= \mathrm{diag}\Big(\Big[-j\pi \Big(-\frac{M_x -1}{2}\Big)  \sin(\phi_1), \cdots,\nonumber\\ & -j\pi \Big(\frac{M_x -1}{2}\Big)  \sin(\phi_1)\Big]\Big),\\
\bD_{i1} =&  \mathrm{diag}\Big(\Big[-j\pi \Big(-\frac{N_x -1}{2}\Big) \sin(\theta_i) \sin(\phi_i), \cdots,\nonumber\\ &-j\pi \Big(\frac{N_x -1}{2}\Big) \sin(\theta_i) \sin(\phi_i)\Big]\Big), \;\text{for}\; i = 2,3, \\
\bD_{i2} =& \mathrm{diag}\Big(\Big[j\pi \Big(-\frac{N_x -1}{2}\Big) \cos(\theta_i) \cos(\phi_i), \cdots,\nonumber\\ &j\pi \Big(\frac{N_x -1}{2}\Big) \cos(\theta_i) \cos(\phi_i)\Big]\Big),\;\text{for}\; i = 2,3,\\ 
\bD_{i3} =& \mathrm{diag}\Big(\Big[-j\pi \Big(-\frac{N_x -1}{2}\Big)  \sin(\phi_i), \cdots, \nonumber \\ &-j\pi \Big(\frac{N_x -1}{2}\Big)  \sin(\phi_i)\Big]\Big), \;\text{for}\; i = 2,3,
\end{align}
under the assumption of half-wavelength inter-element spacing for both the BS and the STAR-RIS UPAs. 

For the above estimators of the channel parameters (unbiased $\hat{\boldsymbol{\nu}}(\by)$), we can calculate the CRLB on the error covariance matrix as follows:
\begin{equation}\label{CRLB_NU}
    \mathbb{E}\{(\boldsymbol{\nu}- \hat{\boldsymbol{\nu}}(\by))(\boldsymbol{\nu}- \hat{\boldsymbol{\nu}}(\by))^\mathsf{H} \} \succeq \bJ^{-1}(\boldsymbol{\nu}),
\end{equation}
where the notation $\bA \succeq \bB$ for square matrices $\bA$ and $\bB$ means $\ba^\mathsf{H} \bA \ba \geq \ba^\mathsf{H} \bB \ba$ for any valid vector $\ba$. 

\subsection{Estimation of 3D Cartesian Coordinates}
We are interested in estimating the 3D Cartesian coordinates of the indoor and outdoor MSs. Therefore, after estimating the channel parameters, we need to map them to 3D Cartesian coordinates, e.g., $\boldsymbol{\kappa} = [x_{\text{U},1}, y_{\text{U},1}, z_{\text{U},1},x_{\text{U},2}, y_{\text{U},2}, z_{\text{U},2}]^\mathsf{T}$, based on the geometrical relationship among the BS, the STAR-RIS, and the two MSs. For the CRLB evaluation of $\boldsymbol{\kappa}$, we resort to the Jacobian matrix $\bT$, which links the connection between the channel parameters $ \boldsymbol{\nu}$ and the 3D Cartesian coordinates of the two MSs $\boldsymbol{\kappa}$. Each $(i,j)$th element of $\bT$ is expressed as:
\begin{equation}
    [\bT]_{ij} = \frac{\partial [\boldsymbol{\nu}]_j}{\partial [\boldsymbol{\kappa}]_i}. 
\end{equation}
We provide the following derivatives for the calculation of the submatrix of the Jacobian matrix related to the outdoor MS: 
\begin{align}\label{theta_x_U_1}
    \partial \theta_i / \partial  x_{\text{U},1} & =- \frac{\sin(\theta_i)}{d_i \cos(\phi_i)},  \\
     \partial \theta_i/ \partial  y_{\text{U},1} & = \frac{\cos(\theta_i)}{d_i \cos(\phi_i)}, \\
     \partial \theta_i / \partial  z_{\text{U},1} & = 0, \\
         \partial \phi_i / \partial  x_{\text{U},1} & =- \frac{\cos(\theta_i)\sin(\phi_i)}{d_i },  \\
     \partial \phi_i/ \partial  y_{\text{U},1} & =- \frac{\sin(\theta_i)\sin(\phi_i)}{d_i }, 
               \end{align}
     \begin{align}
     \partial \phi_i / \partial  z_{\text{U},1} & = \frac{\cos(\phi_i)}{d_i},\\
     \partial d_i / \partial  x_{\text{U},1} & = \cos(\theta_i) \cos(\phi_i),  \\
     \partial d_i/ \partial  y_{\text{U},1} & = \sin(\theta_i) \cos(\phi_i), \\
     \partial d_i / \partial  z_{\text{U},1} & = \sin(\phi_i),\label{d1_z_U_1}
\end{align}
for $i =1$ and $2$. The submatrix of $\bT$ related to the indoor MS can be computed in the same manner. In addition, it can be easily seen that only the channel parameters $\{ \theta_1, \phi_1, d_1,\theta_2, \phi_2, d_2\}$ are related to the coordinates $(x_{\text{U},1}, y_{\text{U},1}, z_{\text{U},1})$ of the outdoor MS, and only the parameters $\{\theta_3, \phi_3, d_3\}$ are related to the coordinates $(x_{\text{U},2}, y_{\text{U},2}, z_{\text{U},2})$ of the indoor MS, as concluded from~\eqref{p_u_1_1} and \eqref{p_u_2}. Therefore, the Jacobian matrix $\bT$ has the following form:
\begin{equation}
\bT = 
\begin{bmatrix}
\bT_{1} & \mathbf{0} \\
\mathbf{0} & \bT_{2} 
\end{bmatrix},
\end{equation} 
where the submatrix $\bT_{1} \in \mathbb{R}^{3\times 6}$ consists of the partial derivatives from~\eqref{theta_x_U_1} to \eqref{d1_z_U_1}, and the submatrix $\bT_{2}\in \mathbb{R}^{3\times 3}$ consists of the partial derivatives related to the indoor MS; this part is omitted here due to the space limitation. 

The Fisher information of $\boldsymbol{\kappa}$ can be then expressed as~\cite{Elzanaty2021} 
\begin{equation}\label{J_kappa}
    \bJ(\boldsymbol{\kappa}) = \bT  \bJ(\boldsymbol{\nu}) \bT^\mathsf{T}. 
\end{equation}
Similar to~\eqref{CRLB_NU}, we have the inequality for the CLRB:
\begin{equation}\label{CRLB_kappa}
    \mathbb{E}\{(\boldsymbol{\kappa}- \hat{\boldsymbol{\kappa}}(\by))(\boldsymbol{\kappa}- \hat{\boldsymbol{\kappa}}(\by))^\mathsf{T} \} \succeq \bJ^{-1}(\boldsymbol{\kappa}).
\end{equation}
The lower bounds on the root mean square error (RMSE) of the position estimation of the outdoor and indoor MSs are:
\begin{align}
 \text{RMSE}_{\text{U},1} = \sqrt{\mathrm{var}(\hat{\bx}_{\text{U},1})} &\geq \sqrt{\mathrm{tr}\{[\bJ^{-1}(\boldsymbol{\kappa})]_{1:3,1:3}\}}, \\
 \text{RMSE}_{\text{U},2} = \sqrt{\mathrm{var}(\hat{\bx}_{\text{U},2})} &\geq \sqrt{\mathrm{tr}\{[\bJ^{-1}(\boldsymbol{\kappa})]_{4:6,4:6}\}}.
\end{align}

\subsection{Localization-Optimal Design for the STAR-RIS}\label{Optimal_Design_of_STAR_RIS}
By introducing $\bG_1 \triangleq [\frac{\partial \boldsymbol{\mu}} {\partial \theta_1}, \frac{\partial \boldsymbol{\mu}} {\partial \phi_1}, \frac{\partial \boldsymbol{\mu}} {\partial d_1}, \frac{\partial \boldsymbol{\mu}} {\partial \theta_2}, \frac{\partial \boldsymbol{\mu}} {\partial \phi_2}, \frac{\partial \boldsymbol{\mu}} {\partial d_2} ]$, $\hat{\bG}_1 \triangleq \bG_1 \bT_{1}^\mathsf{H}$, $\bG_2 \triangleq [\frac{\partial \boldsymbol{\mu}} {\partial \theta_3}, \frac{\partial \boldsymbol{\mu}} {\partial \phi_3}, \frac{\partial \boldsymbol{\mu}} {\partial d_3}]$, and $\hat{\bG}_2 \triangleq \bG_2 \bT_{2}^\mathsf{H}$, the expression of $\bJ^{-1}(\boldsymbol{\kappa})$ in~\eqref{CRLB_NU} can be expressed as~\cite{scharf1993geometry, Pakrooh2015}

\begin{equation}\label{bJ_inv}
      \bJ^{-1}\!(\boldsymbol{\kappa}) \!=\! \frac{\sigma^2}{P}
      \begin{bmatrix}
     (\hat{\bG}_1^\mathsf{H} (\bI \!-\! \bP_{\hat{\bG}_2})\hat{\bG}_1)^{-1} & * \\
      * &  (\hat{\bG}_2^\mathsf{H} (\bI \!-\! \bP_{\hat{\bG}_1})\hat{\bG}_2 )^{-1}
      \end{bmatrix}\!\!,
\end{equation}
where $\bP_{\hat{\bG}_i} = \hat{\bG}_i( \hat{\bG}_i^\mathsf{H} \hat{\bG}_i)^{-1}\hat{\bG}_i^\mathsf{H} $ is the orthogonal projection onto the column space of $\hat{\bG}_i$ for $i = 1$ and $2$.

In order to simplify the diagonal terms in~\eqref{bJ_inv}, we rewrite $\bG_1$ and $\bG_2$ as $\bG_1 = [\eta_1\bA_1\bH_1, \; \eta_1 \epsilon_2 \bA_2  \bH_2]$ and $\bG_2 = \eta_2 \epsilon_1 \bA_3\bH_3$, where $\bH_1 = [(\bD_{11} \otimes \bI_{M_x})\bh_1, \; (\bD_{12} \otimes \bI_{M_x}  + \bI_{M_z} \otimes \bD_{13})\bh_1, \; \frac{-j2\pi d_1/\lambda -1}{d_1} \bh_1]$, $\bH_2 = [(\bD_{21} \otimes \bI_{N_x})\bh_2, \; (\bD_{22} \otimes \bI_{N_x}  + \bI_{N_z} \otimes \bD_{23})\bh_2, \; \frac{-j2\pi d_2/\lambda -1}{d_2} \bh_2]$, and $\bH_3 = [(\bD_{31} \otimes \bI_{N_x})\bh_3, \; (\bD_{32} \otimes \bI_{N_x}  + \bI_{N_z} \otimes \bD_{33})\bh_3, \; \frac{-j2\pi d_3/\lambda -1}{d_3} \bh_3]$. It can be seen that $\bA_2$ is a function of $\bar{\boldsymbol{\Omega}}_2$ and $\bA_3$ is a function of $\bar{\boldsymbol{\Omega}}_1$, while $\bA_1$ is independent of $\bar{\boldsymbol{\Omega}}_1$ and $\bar{\boldsymbol{\Omega}}_2$. By dividing $\bT_{1}$ into two submatrices, i.e., as $\bT_{1} = [\tilde{\bT}_{1}\in \mathbb{R}^{3\times 3}, \bar{\bT}_{1}\in \mathbb{R}^{3\times 3}]$, we can derive the expressions $\hat{\bG}_1 = \eta_1\bA_1\bH_1\tilde{\bT}_{1}^\mathsf{H} +  \eta_1 \epsilon_2 \bA_2  \bH_2\bar{\bT}_{1}^\mathsf{H}$ and $\hat{\bG}_2 = \eta_2 \epsilon_1 \bA_3\bH_3\bT_{2}^\mathsf{H}$.

For the sake of tractability for the STAR-RIS optimization (i.e., $\bar{\boldsymbol{\Omega}}_1$ and $\bar{\boldsymbol{\Omega}}_2$), we maximize the principal angle (within $[0,\pi/2]$) between the subspaces of $\hat{\bG}_1$ and $\hat{\bG}_2$ by following~\cite{scharf1993geometry}. The optimal solutions are founded when $ [\mathbf{1}, \bar{\boldsymbol{\Omega}}_2^\mathsf{T}]$ is orthogonal to $\bar{\boldsymbol{\Omega}}_1^\mathsf{T}$. In this regard, the largest principal angle, e.g., $\pi/2$, is obtained~\cite{golub2013matrix}. The choices for the $\bar{\boldsymbol{\Omega}}_1$ and $\bar{\boldsymbol{\Omega}}_2$ can be the non-overlapping parts of a DFT or a Hadamard matrix.\footnote{Hadamard matrices possess good properties, since they only contain $\{-1, 1\}$. Therefore, only $1$-bit quantization is needed for both the reflection and refraction matrices at the STAR-RIS.} Note that we need to assume that $K \geq 2 N$ here in order to guarantee that both $\bar{\boldsymbol{\Omega}}_1$ and $\bar{\boldsymbol{\Omega}}_2$ are optimal.  

\section{Numerical Results}
In this section's numerical investigation, we have set the system parameters as follows: $\bp_\text{B} =(0, 0, 8)^\mathsf{T}$, $\bp_\text{R} =(2, 2, 5)^\mathsf{T}$, $\bp_{\text{U},1} =(5, 1, 2)^\mathsf{T}$, and $\bp_{\text{U},2} =(1, 5, 2)^\mathsf{T}$. The numbers of BS antennas, STAR-RIS elements, and SRSs from each MS were set as $M = 16$, $N = 64$, and $K = 128$. The signal-to-noise ratio (SNR) is defined as $P/\sigma^2$. 


\subsection{Results on the Channel Parameters' Estimation}
\begin{figure}[t]
	\centering
\includegraphics[width=0.99\linewidth]{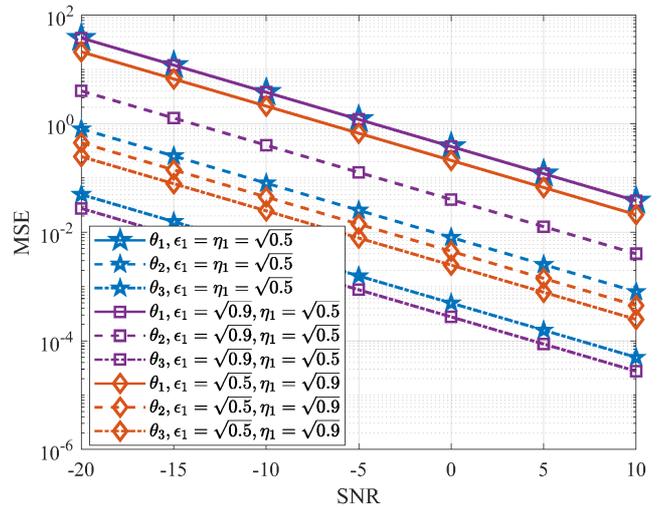}
	\caption{CRLB on the estimation of $\{\theta_i\}$ for $i =1,2$ and $3$ with different pairs of $\epsilon_1$ and $\eta_1$. }
		\label{MSE_Theta_effect_eta_epsilon}
\end{figure}

\begin{figure}[t]
	\centering
\includegraphics[width=0.95\linewidth]{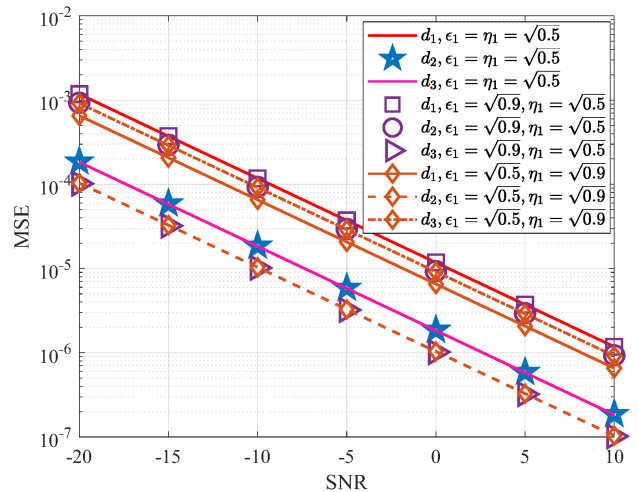}
	\caption{CRLB on the estimation of $\{d_i\}$ for $i =1,2$ and $3$ with different pairs of $\epsilon_1$ and $\eta_1$. }
		\label{MSE_d_effect_eta_epsilon}
\end{figure}

We evaluate the estimation of the intermediate parameters with different setups for the STAR-RIS energy splitting coefficient $\epsilon_1$ and the power allocation coefficient $\eta_1$, i.e., $\epsilon_1 = \eta_1 = \sqrt{0.5}$, $\epsilon_1 = \sqrt{0.9}, \eta_1 = \sqrt{0.5}$, and $\epsilon_1 = \sqrt{0.5}, \eta_1 = \sqrt{0.9}$. The simulation results on estimation of $\theta_i$'s and $d_i$'s are shown in Figs.~\ref{MSE_Theta_effect_eta_epsilon} and~\ref{MSE_d_effect_eta_epsilon} in terms of the mean square error (MSE). Due to the similar estimation performance for $\theta_i$'s, the results for $\phi_i$'s are omitted here. From the simulation results, we can see that when $\epsilon_1$ increases with fixed $\eta_1$, better performance on the estimation of parameters in $\bh_3$ can be achieved. Meanwhile, the performance degrades for the parameter estimation in $\bh_1$ and $\bh_2$ when $\eta_1$ decreases (less transmission power at the outdoor MS). Note that when fixing $\eta_1$, the performance for the parameter estimation in $\bh_1$ will not change with $\epsilon_1$. 

\subsection{CRLBs on 3D Localization}
\begin{figure}[t]
	\centering
\includegraphics[width=0.95\linewidth]{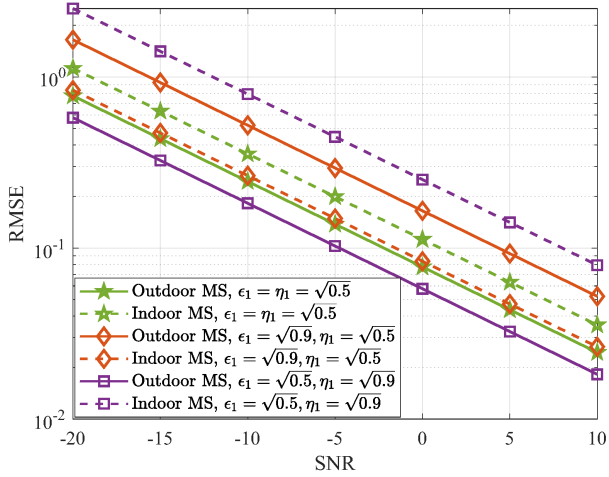}
	\caption{CRLB on the estimation of the 3D Cartesian coordinates of the two MSs with different pairs of $\epsilon_1$ and $\eta_1$. }
		\label{RMSE_3D_Loc_effect_eta_epsilon}
		\vspace{-0.5cm}
\end{figure}

Based on the expressions~\eqref{J_kappa} and \eqref{CRLB_kappa}, we can calculate the CRLBs for the estimation of the 3D Cartesian coordinates of the indoor and outdoor MSs. The simulation results are shown in Fig.~\ref{RMSE_3D_Loc_effect_eta_epsilon} in terms of RMSE. The performance of 3D localization has been controlled by the two parameters $\epsilon_1$ and $\eta_1$. When $\epsilon_1 = \eta_1 =\sqrt{0.5}$, the performance gap between the two MSs' position estimation is small. However, in the other two cases, the gap is obvious. In this sense, well-selected $\epsilon_1$ and $\eta_1$ can satisfy the quality of services (QoSs) and user fairness for both MSs simultaneously. 

\begin{figure}[t]
	\centering
\includegraphics[width=0.99\linewidth]{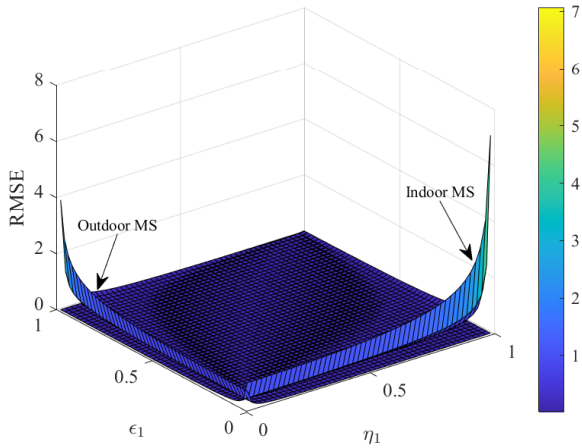}
	\caption{The effect of $\eta_1$ and $\epsilon_1$ on the 3D localization, when the SNR is fixed to $15$ dB. }
		\label{heat_map}
 		\vspace{-0.5cm}
\end{figure}

More comprehensive results on $\epsilon_1$ and $\eta_1$ on the 3D localization are shown in Fig.~\ref{heat_map}. In general, a small $\epsilon_1$ and a large $\eta_1$ offer poor performance for the indoor MS. On the contrary, a large $\epsilon_1$ and a small $\eta_1$ results in poor performance for the outdoor MS. However, for most of the cases, where it holds $\eta_1,\epsilon_1 >0.3$, we can achieve promising 3D localization in term of the RMSE for both MSs.

 \subsection{Effect of the STAR-RIS Design}\label{STAR_RIS_Design}
\begin{figure}[t]
	\centering
\includegraphics[width=0.99\linewidth]{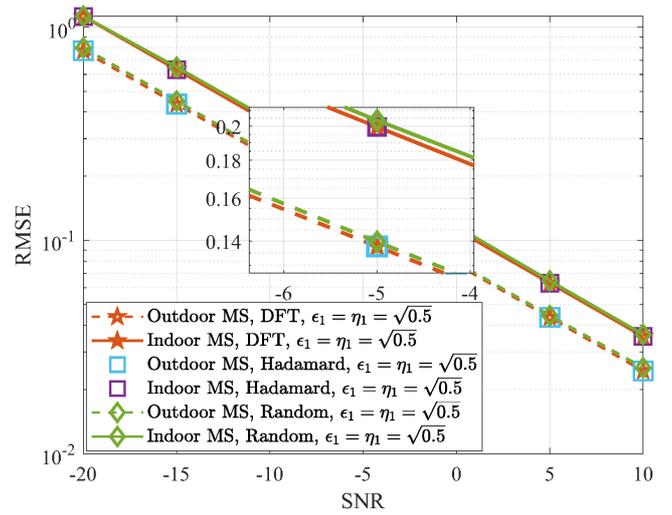}
	\caption{The effect of the STAR-RIS design on the 3D localization. }
		\label{Effect_of_STAR_RIS_Design}
 		\vspace{-0.5cm}
\end{figure}

We evaluate the effect of the STAR-RIS design by considering three different cases: i) $\bar{\boldsymbol{\Omega}}_1$ and $\bar{\boldsymbol{\Omega}}_2$ are part of a DFT matrix; ii) $\bar{\boldsymbol{\Omega}}_1$ and $\bar{\boldsymbol{\Omega}}_2$ are part of a Hadamard matrix; and iii) the phases of  $\bar{\boldsymbol{\Omega}}_1$ and $\bar{\boldsymbol{\Omega}}_2$ are randomly generated. The simulation results are shown in Fig.~\ref{Effect_of_STAR_RIS_Design}. We observe that the first two cases have the same performance and outperform the third one, since according to Section~\ref{Optimal_Design_of_STAR_RIS} they are optimal. The performance gain is not obvious due to the fact that, for large $K$ and $N$, case iii) approximates the optimal solutions (i.e., the subspaces of $\hat{\bG}_1$ and $\hat{\bG}_2$ are nearly orthogonal).
 
\section{Conclusion and Future Work}
In this paper, we studied the fundamental 3D localization performance limits of STAR-RIS-aided mmWave MIMO systems for simultaneously serving one indoor MS and one outdoor MS. We also investigated the effect of energy splitting at the STAR-RIS and the power allocation between the two MSs to offer some useful insights for practical implementations. Moreover, the optimal design of the STAR-RIS reflection and refraction configuration matrices was derived by maximizing the principal angle of two associated subspaces.

In the future, we will focus on the design of practical localization algorithms, which attain this paper's theoretical performance limits. We will also extend the CRLB analysis to general multipath scenarios and consider arbitrary orientations between the UPAs of the BS and the STAR-RIS.

\bibliographystyle{IEEEtran}
\bibliography{IEEEabrv,Ref}

\end{document}